\documentclass{PoS}
\pdfoutput=1

\title{Emission line profiles and X-ray observations of Broad and
 Narrow Line Seyfert 1 Galaxies}

\ShortTitle{Emission line and X-ray properties of Broad and Narrow Line Seyfert 1 galaxies}

\author{\speaker{G. La Mura} \\ 
Department of Astronomy - University of Padua, Italy \\
E-mail: \email{giovanni.lamura@unipd.it}}

\author{S. Ciroi \\
Department of Astronomy - University of Padua, Italy \\
E-mail: \email{stefano.ciroi@unipd.it}}

\author{V. Cracco \\
Department of Astronomy - University of Padua, Italy \\
E-mail: \email{valentina.cracco@unipd.it}}

\author{D. Ili\'c \\
Department of Astronomy - Faculty of Mathematics - Unviversity of Belgrade, Republic of Serbia \\
E-mail: \email{dilic@matf.bg.ac.rs}}

\author{L. \v C. Popovi\'c \\
Astronomical Observatory of Belgrade, Republic of Serbia \\
E-mail: \email{lpopovic@aob.rs}}

\author{P. Rafanelli \\
Department of Astronomy - University of Padua, Italy \\
E-mail: \email{piero.rafanelli@unipd.it}}

\abstract{Narrow Line Seyfert 1 galaxies represent a distinct category in the larger family of Type 1 Active Galactic Nuclei. The occurrence of emission line profiles not broader than 2000$\,{\rm km\,s}^{-1}$, combined with luminosity estimates comparable to those of their broad line emitting analogues, suggests that some extreme property is characterizing the fueling of their central engines. Two hypotheses are more commonly considered: on the one hand, it was suggested that the peculiar features of NLS1s might be interpreted as the consequence of a nearly pole-on view over a geometrically flattened source; on the other, there are arguments favoring the interpretation of a relatively low mass black hole, accreting at very high rates. Both explanations provide advantages and drawbacks, but, in spite of the different assumed physics, they agree in identifying NLS1s as the crucial test ground for our understanding of the Broad Line Region structure and dynamics in AGN. Here we report the results obtained investigating asynchronous optical and X-ray spectroscopic observations, respectively extracted from the Sloan Digital Sky Survey and the XMM-Newton science archives, of a sample of both broad and narrow line emitting Seyfert 1 galaxies. Exploiting data collected by the different instruments carried on board XMM, we try to identify the various components which sum up in the observed broad band spectra of the two classes. We discuss the relation that thermal, non-thermal, line emission and broad band absorption components of the X-ray source show with the optical emission line profiles, as interpreted in the framework of a composite Broad Emission Line Region structural model.
}

\FullConference{Narrow-Line Seyfert 1 Galaxies and their place in the
  Universe - NLS1,\\ April 04-06, 2011\\ Milan Italy
}

\begin{document}
\newcommand{\ion}[2]{#1~{\small #2}}
\newcommand{\de}{\rm d}

\section{Introduction}
After more than 25 years of extensive investigations, Narrow Line Seyfert 1 (NLS1) galaxies still pose some of the most intriguing challenges to our understanding of Active Galactic Nuclei (AGN). Following their first detections and the subsequent identification as a peculiar Seyfert class \cite{Davidson78, Osterbrock85}, it was quickly realized that the physics of nuclear activity is somehow carrying out its extreme configurations in these objects. Although they were first noted to have the characteristic features of Type 1 Seyfert spectra, but with recombination lines only slightly broader than forbidden ones, subsequent studies pointed out many peculiar properties that extend well beyond a mere line width based distinction.

Applying the correlation analysis of Boroson \& Green \cite{Boroson92} among the emission line profiles and intensities, NLS1 galaxies were found to posses the smallest values of the [\ion{O}{III}] / H$\beta$ ratio and the strongest \ion{Fe}{II} multiplets observed in Seyfert 1s \cite{Sulentic00, Sulentic02, Boroson02}. Significant line shifts and asymmetries were detected in \ion{C}{IV} \cite{Wills00, Leighly04, Sulentic07} and [\ion{O}{III}] \cite{Zamanov02, Bian05, Boroson05}, while a systematically high [\ion{S}{II}] 6716/6731 intensity ratio indicates a low density Narrow Line Region (NLR) environment \cite{Xu07}. At X-ray energies, the spectra are usually characterized by a steep power-law continuum in the hard band (2 - 10~keV) and by an enhanced variability with respect to broad line Seyfert 1s (e. g. \cite{Leighly99, Grupe04, McHardy06}). The optical variability, on the contrary, is not exceptional \cite{Giannuzzo98, Klimek06, Ai10}. Komossa et al. \cite{Komossa06} pointed out that NLS1s are rarely radio-loud, though signatures of jet activity have sometimes been spotted \cite{Grupe00, Abdo09a, Abdo09b}.

Several interpretations have been proposed, in order to explain the characteristic features mentioned above. Geometric effects, such as flattening and inclination, as well as winds and density properties, metallicity, absorptions, and high accretion rates onto relatively low mass black holes were taken into account (see \cite{Komossa08} for a comprehensive review). At present, the most widely accepted paradigm assumes that NLS1 properties mainly descend from the small mass of the black hole powering the central engine, estimated to be in the range of $10^6 - 10^7$M$_\odot$, against the typical masses of $10^8 - 10^9$M$_\odot$ found in broad line Seyfert 1s (BLS1s) and QSOs. Since both NLS1 and BLS1 nuclei span the same range in luminosity \cite{Peterson00}, it follows that the accretion rates of NLS1s are considerably higher than in BLS1s. This interpretation agrees with the observed relationship among the black hole mass and the luminosity of the host galaxy bulge, which is generally faint in NLS1s \cite{Botte04}, but it raised some problems in the corresponding relationship associated with the stellar velocity dispersion. Due to the difficulties in getting accurate determinations of stellar kinematics in AGNs, however, the velocity dispersion of the narrow line region gas has been sometimes used as a surrogate. Botte et al. \cite{Botte05} questioned the validity of such a proxy, pointing out that additional contributions to the gas kinematics, driven by winds associated to the high accretion rates of the central engine, could originate the discrepancy, which, instead, is appreciably reduced when direct measurements of the stellar velocity dispersion are available.

As a contribution to the ongoing investigation on the nature of NLS1 galaxies, here we present a compared analysis of the optical and X-ray spectra of BLS1s and NLS1s. We study the role played by geometric effects and intrinsic properties of the central engine in the formation of the observed emission line profiles and their influence on the ionization conditions of the line emitting plasma. For this reason we shall organize our discussion as follows: in \S2 we present the relationship among broad line plasma physics and the observed emission line profiles; in \S3 we describe the observations taken into account in our analysis, while \S4 provides an overview of our results and the final discussion.

\section{Plasma diagnostics with Boltzmann Plots}
Type 1 AGN spectra are fundamentally characterized by the presence of a strong non-stellar continuum and by several broad and narrow emission lines. While it is observed that a narrow component appears in every emission line, the broad components, on the contrary, mostly affect the permitted and some inter-combination lines. It is, therefore, clear that the origin of line emission can be traced back to different regions, having distinct physical properties. In the so called Narrow Line Region (NLR), which is located at a large distance from the central continuum source, so that the emission lines have profiles with FWHM~$< 10^3\,{\rm km\,s}^{-1}$, we find a low density plasma ($N_e \approx 10^6\,{\rm cm}^{-3}$), where the meta-stable levels of ion species, which are charged by collisional processes, can survive long enough to fall back to their ground configuration with spontaneous radiative decays. Deep in the gravitational potential of the central engine, much closer to the continuum source (at a distance $R \leq 1\,$pc), instead, the Broad Line Region (BLR) hosts a high density medium ($10^8\,{\rm cm}^{-3} \leq N_e \leq 10^{13}\,{\rm cm}^{-3}$), where forbidden lines are suppressed by the collisional de-excitation of their upper levels. Only the emission lines with high transition probabilities can originate in the BLR and the kinematics of the line emitting gas results in a profile broadening with FWHM~$> 10^3\,{\rm km\,s}^{-1}$.

The suppression of forbidden lines, as a consequence of the significant influence of collisional processes, makes it difficult to estimate the BLR physical properties, by means of the usual diagnostic techniques applied to other nebular environments. However, if we consider an emission line, originated by the transition from an upper level $u$ to a lower level $l$, we can express the associated intensity as:
$$I_{u l} = \frac{h c}{\lambda_{u l}} A_{u l} \int_{R_1}^{R_2} N_u(r)\cdot e^{-\tau(r)} \de r, \eqno(1)$$
where $A_{u l}$ is the probability of a spontaneous radiative decay, $N_u(r)$ the number density of the emitting ions, $\tau(r)$ is the optical depth of the layer, $h$, $c$, and $\lambda_{u l}$ represent the Planck constant, the speed of light, and the line wavelength, while $R_1$ and $R_2$ are the boundaries of the emitting region. In the case of an optically thin plasma, with the population of the high excitation energy levels following the Saha-Boltzmann distribution (plasma in a {\it Partial Local Thermodynamic Equilibrium}, or PLTE; see \cite{Popovic03, Popovic06}), Eq.~(1) takes the form of:
$$I_{u l} = \frac{h c}{\lambda_{u l}} A_{u l} g_u N_0 \ell e^{-E_{u 0} / k_B T}, \eqno(2)$$
in which $g_u$ is the statistical weight of the upper level, $\ell$ is the spatial extension of the emitting region, $N_0$ is the number density of the ions in their unexcited configuration, while $E_{u 0}$, $k_B$, and $T$ are the excitation energy of the level, the Boltzmann constant, and the plasma electron temperature, respectively. Introducing a normalized line intensity, with respect to the atomic constants that characterize the transitions, we find from Eq.~(2):
$$I_n = \frac{\lambda_{u l} I_{u l}}{g_u A_{u l}} = h c N_0 \ell e^{-E_{u 0} / k_B T}. \eqno(3)$$
If we take into account a series of transitions originated by a particular ion species, the normalized intensity of each line is a function of the upper level energy, that we can express as:
$$\log I_n = -\frac{\log e}{k_B T} E_{u 0} + const., \eqno(4)$$
where it has been assumed that the line emitting region is the same for the whole transition series.

\begin{figure}[t]
\begin{center}
\includegraphics[width=12cm]{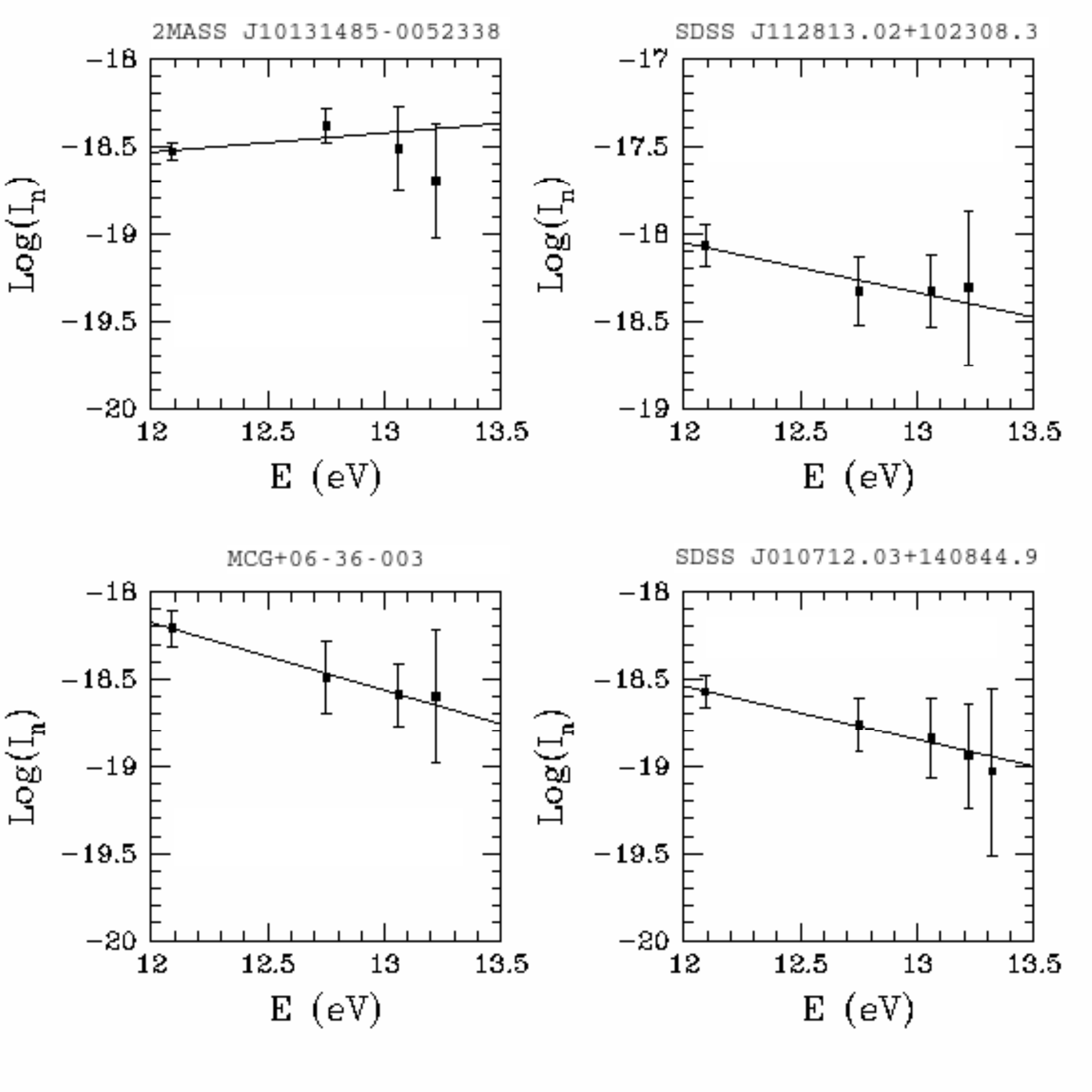}
\end{center}
\caption{Four examples of Boltzmann Plot applied to the broad Balmer emission line components: (upper left) no linear fit to the Balmer series; (upper right) poor fit to the expected linear behaviour; (lower left) the BP provides an appreciable fit to the Balmer lines up to H$\delta$; (lower right) the broad components of the Balmer series are detected up to H$\epsilon$ and they fit in the linear function. \label{f01}}
\end{figure}
Eq.~(4) defines the Boltzmann Plot (BP) of the transition series, which, in the framework of the assumptions described above, predicts the normalized line intensities to behave as a specific function of their excitation energies and of the plasma electron temperature. The BP method, therefore, represents a valuable tool to investigate the physical conditions in a line emitting plasma, where collisional processes are frequent enough to affect the population of excited ion species. In La Mura et al. \cite{LaMura07}, this idea was explored with the application of the technique to a statistically relevant sample of type 1 AGNs, extracted from the 3$^{rd}$ data release of the {\it Sloan Digital Sky Survey} (SDSS). As it is illustrated in Fig.~\ref{f01}, where some examples of BP applied to the broad components of the Balmer lines are given, the technique led to the identification of four main possibilities, corresponding to different physical scenarios in the line emitting plasma. 

\begin{figure}[t]
\begin{center}
\includegraphics[width=12cm]{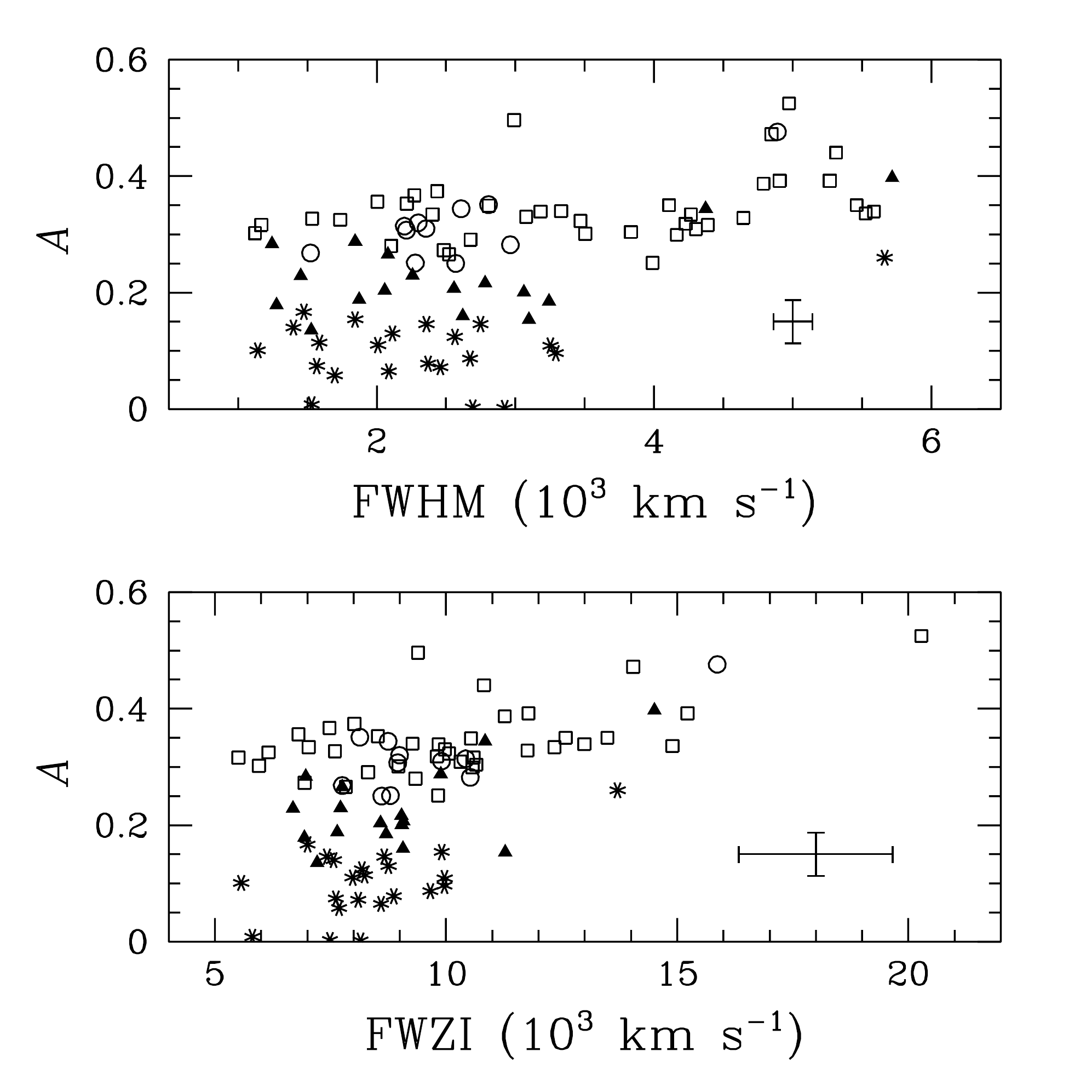}
\end{center}
\caption{The temperature parameter $A$ plotted vs. the FWHM (upper panel) and FWZI (lower panel) of broad emission line components for the different BP results: asterisks are objects with no reasonable fit to Eq.~(4); filled triangles are objects with a poor linear fit; open squares represent sources with a good fit and the broad emission line components detected up to H$\delta$; open circles are objects  featuring a good fit and the broad emission line component up to H$\epsilon$. Data from \cite{LaMura07}. \label{f02}}
\end{figure}
In general, it is observed that a satisfactory match with the underlying assumptions is achieved in approximately 30\%\ of the selected objects, with an improved statistics in the domain of sources with very broad emission lines. This result is illustrated in Fig.~\ref{f02}, where the temperature parameter $A = -\log e / k_B T$, corresponding to the slope of the linear function defined in Eq.~(4), is compared to the width of the broad emission line components. In the range of narrow line emitting objects, instead, the BP finds indications of a higher degree of plasma ionization, probably due to a stronger interaction with the ionizing radiation field, which leads to the gradual loss of the equilibrium conditions required by the method. Furthermore, in the objects where the fundamental BP assumptions are still appreciably holding, the inferred temperatures are averagely higher than the values found in broad line emitting sources.

\section{Data description and analysis}
The detection of larger temperatures or, more generally, of an increasing degree of ionization in the BLR plasma of objects with narrow emission line profiles is an intriguing point, suggesting that the characteristics of the ionizing radiation observed in NLS1 galaxies are different with respect to their broad line emitting counterparts. Such differences can be naturally expected as a direct consequence of the high accretion rates required by the interpretation of NLS1s as low mass objects. However, the gradual loss of fundamental BP assumptions in the range of strongly photoionized plasmas forces some care in the interpretation of the results. An independent test of the observations involves the study of type 1 AGN spectra over a broader range of frequencies.

\begin{table}[t]
\begin{center}
\caption{List of targets selected from the combined optical and X-ray observation sample providing the object names, coordinates (J2000), measured X-ray fluxes, and redshifts. \label{t01}}
\begin{tabular}{lcccc}
\hline
\hline
Name & R. A. & Dec. & $F_X$ (0.2~keV - 12~keV) & z \\
 & (h:m:s) & (deg:m:s) & ($10^{-13}\, {\rm erg\, cm^{-2}\, s}^{-1}$) & \\
\hline
2MASX J03063958+0003426 & 03:06:39.58 & +00:03:43.2 & 17.31 $\pm$ 0.18 & 0.107 \\
MCG+04-22-042 & 09:23:43.01 & +22:54:32.4 & 348.93 $\pm$ 2.02 & 0.033 \\
Mrk 110 & 09:25:12.87 & +52:17:10.5 & 593.08 $\pm$ 1.55 & 0.035 \\
PG 1114+445 & 11:17:06.39 & +44:13:33.3 & 35.91 $\pm$ 0.29 & 0.144 \\
PG 1115+407 & 11:18:30.30 & +40:25:54.1 & 47.71 $\pm$ 0.34 & 0.155 \\ 
2E1216.9+0700 & 12:19:30.87 & +06:43:34.8 & 18.61 $\pm$ 0.42 & 0.081 \\
Mrk 50 & 12:23:24.14 & +02:40:44.4 & 278.17 $\pm$ 4.70 & 0.023 \\
Was 61 & 12:42:10.60 & +33:17:03.0 & 131.11 $\pm$ 0.29 & 0.044 \\
PG 1352+183 & 13:54:35.69 & +18:05:17.5 & 55.99 $\pm$ 0.55 & 0.151 \\
Mrk 464 & 13:55:53.52 & +38:34:28.7 & 47.17 $\pm$ 0.75 & 0.051 \\
PG 1415+451 & 14:17:00.83 & +44:56:06.3 & 33.49 $\pm$ 0.26 & 0.114 \\
NGC 5683 & 14:34:52.48 & +48:39:42.9 & 118.25 $\pm$ 1.50 & 0.036 \\
Mrk 290 & 15:35:52.42 & +57:54:09.5 & 173.49 $\pm$ 0.71 & 0.030 \\
Mrk 493 & 15:59:09.67 & +35:01:47.3 & 140.22 $\pm$ 0.69 & 0.031 \\
\hline
\end{tabular}
\end{center}
\end{table}
Here we report on the analysis of a sample of objects, for which high quality spectroscopic information could be retrieved by matching the SDSS Data Release 7 database with the XMM Newton observation archive. A selection of type 1 AGN optical spectra, from objects corresponding to X-ray sources brighter than $5 \cdot 10^{-13} {\rm erg\, cm^{-2}\, s^{-1}}$ in the 0.2~keV - 12~keV energy range at the time of observation (a flux limit based on the quality of observations with the XMM instruments in exposure times of a few $10^3\,$s) led to the identification of 122 potential targets, having both broad and narrow emission lines. Work on this sample is still in progress, but Table~\ref{t01} provides a list of the sources that have been studied so far. In the following we describe the data analysis and the information extracted from the optical and X-ray spectra.

\subsection{Optical spectra}
The identification of the BLR component in the spectra of type 1 AGN requires a careful data handling process, which must account for the several contributions that overlap on the nuclear source in the observation. A great advantage of the SDSS database is that it provides a standard calibration of data in physical units, therefore largely simplifying the reduction process. The fixed fiber spectrograph aperture, on the other hand, implies that various fractions of the host galaxy light are recorded, together with the AGN signal, depending on the target redshift. Moreover, extinction from the inter stellar medium (ISM), arising in our own Galaxy, must also be taken into account.

\begin{figure}[t]
\begin{center}
\includegraphics[width=\textwidth]{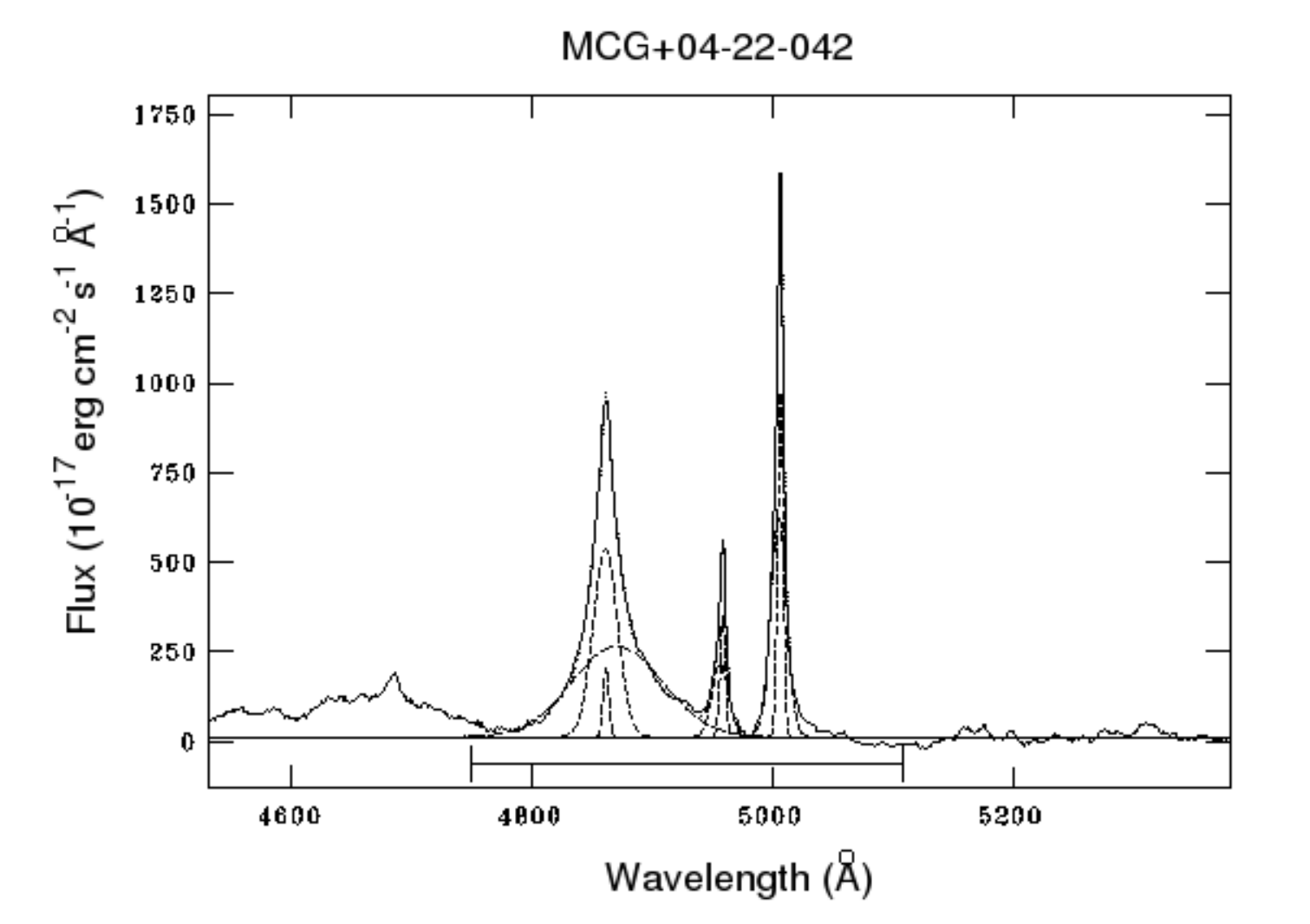}
\caption{Multiple gaussian decomposition of H$\beta$ and [\ion{O}{III}] $\lambda\lambda$ 4959,5007 from the spectrum of MCG+04-22-042. The subtraction of the NLR contributions is performed assuming the well defined [\ion{O}{III}] $\lambda$ 5007 as a template for the profiles of narrow emission lines. \label{f03}}
\end{center}
\end{figure}
The preliminary reduction process, producing the spectra of our selected targets for subsequent analysis, involved the Galactic extinction correction and the subtraction of the signals arising from outside the BLR itself. At first, we applied the correction for Galactic extinction, using a continuous extinction curve, in the form of the function proposed by Cardelli et al. \cite{Cardelli89}, deriving the absorption coefficient $A_V$ in the direction of observation \cite{Schlegel98}. After the corrected spectra were taken to their rest frame, using the peak of the [\ion{O}{III}] $\lambda$5007 as a reference, we proceeded with the removal of the underlying continuum and of the emission line blends, affecting the profiles of the broad spectral features. An example of this final step is shown in Fig.~\ref{f03}, where the profiles of the emission lines in the spectral region of H$\beta$ are modeled by means of multiple gaussian fits. In this technique, we assume that the narrow emission lines can be represented by a standard template profile, built to match the [\ion{O}{III}] $\lambda$5007 spectral feature and scaled to the appropriate intensity, and we use multiple gaussian components to fit the broad contributions. The result is a fair subtraction of the NLR spectrum from the observation, but the isolated BLR component still retains some relevant concerns. Indeed, it has been pointed out that the broad emission lines are often blended together, affected by the contamination of the \ion{Fe}{II} multiplets, and probably complex in their own origin.

In this report we shall not deal with the complete treatment of the techniques that are required to properly account for the complexity of the broad line profiles and for the problem related to the \ion{Fe}{II} multiplets, but we note that, while the former is usually a matter of debate in the domain of broad line emitting sources, the latter is of particular importance for NLS1s. Here we limit our analysis to the identification of those contributions that are most likely arising as an emission within the BLR and to the measurement of their fundamental properties. The interested reader may find more details on these topics in the literature (see e. g. \cite{LaMura09, Kovacevic10}).

\subsection{X-ray spectra}
In the X-ray domain, the energy resolution of our instruments is in most cases still too low for a proper detection of spectral line features. High dispersion X-ray spectroscopy in the low energy band ($0.1\, {\rm keV} \leq E \leq 2.0\, {\rm keV}$) is possible in the case of bright objects, but it requires large amounts of observational time to provide significant data. Broad band spectroscopy in the energy range $0.2\, {\rm keV} \leq E \leq 12.0\, {\rm keV}$, on the other hand, is much more affordable, although the interpretation of results relies onto a different approach, with respect to the classical spectral analysis. Basically it is assumed that the observed spectrum can be regarded as the combination of a set of fundamental components. These build a global model, which, after the instrument characteristics have been taken into account, is used to fit the observation and to estimate the identification likelihood of every component.

The X-ray data considered in our work have been collected with the MOS1, MOS2, and pn detectors of the EPIC camera carried by the XMM observatory. The calibration of observations was accomplished by means of the standard procedures of the XMM Science Analysis Software (SAS version 10.0.0), using the \texttt{emproc} and the \texttt{epproc} tasks to extract observations from
the data packages, then the \texttt{evselect} and \texttt{tabgtigen} to filter the observations from spurious signals and background effects. We collected our source and background spectra in circular regions of 30'' in radius and we used the \texttt{rmfgen} and \texttt{arfgen} tasks to derive the instrument response information. Finally, the spectra were binned to have at least 15 counts per channel.

\begin{figure}[t]
\begin{center}
\includegraphics[width=\textwidth]{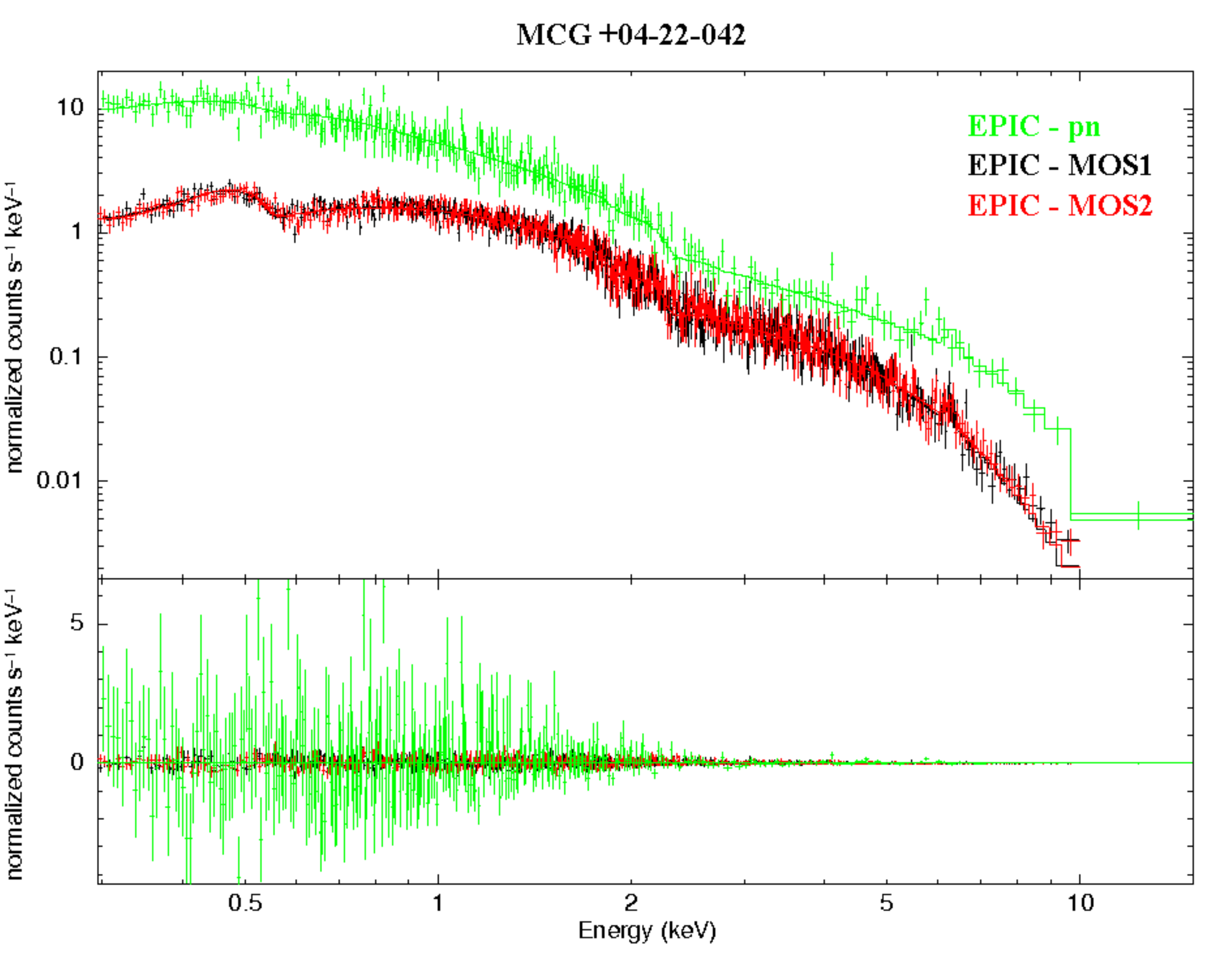}
\caption{X-ray spectra of MCG+04-22-042 recorded by the MOS1 (black), MOS2 (red), and pn (green) detectors of the EPIC camera. The continuous lines represent the model that is assumed to simultaneously fit the observations, producing the residual pattern illustrated in the bottom panel. \label{f04}}
\end{center}
\end{figure}
Our choice to use the different detectors of the EPIC camera results in the extraction of approximately simultaneous observations of the targets, performed by instruments with different energy responses. This circumstance, when available, took us to exploit the technique of simultaneous modeling of the spectra, where it is required that a single model fits distinct data sets. We give an example of this technique in Fig.~\ref{f04}. To build the models, we used the XSPEC software (version 12.6.0q), fitting the spectra in several subsequent steps. At first, we ignored the low energy band data and we reproduced the spectrum between 2~keV and 10~keV by means of a simple power law, combined to a gaussian function at a reference energy of 6.4~keV, accounting for the Fe~K$\alpha$ emission line. As a second step, we froze the power law component and we included in the fit the low energy band between 0.3~keV and 2~keV. This is a particularly complex region of the spectrum, where the contribution of unresolved emission and absorption features from highly ionized species can be relevant \cite{Boller07}. Since the low spectral resolution and the noise fluctuations did not give us the opportunity to significantly distinguish among the possible theoretical models that have been developed to account for the spectral energy distribution in this band, we added a combination of thermal components, using up to two contributions, depending on the resulting improvement of the fit. Once every component had a steady role in subsequent iterations, we allowed for the fit of the whole spectrum. The final model was required to account for the X-ray absorption expected from the neutral gas column density along the observation direction within our Galaxy.

In order to estimate the reliability of every component, we only accepted additional contributions leading to significant improvements of the fit quality, for which we required a reduced $\chi^2 \approx 1$ and a probability of getting such residuals from a correct model not below 30\%. Uncertainty estimates on the model parameters were computed at a 95\% confidence range by applying the error analysis based on the exploration of the parameter space about the best fit model.

\begin{table}[t]
\caption{Spectral properties measured in the optical and X-ray domains of the considered sample. For every object, we provide the FWHM of H$\beta$, the central energy $E_{Fe}$ of the Fe emission, the equivalent width $EW_{Fe}$, and the power law photon index $\Gamma$, estimated in the 2~keV - 10~keV energy band. \label{t02}}
\begin{center}
\begin{tabular}{lcccc}
\hline
\hline
Name & FWHM$_{{\rm H}\beta}$ & $E_{Fe}$ & $EW_{Fe}$ & $\Gamma$\\
 & (km s$^{-1}$) & (keV) & (eV) & \\
 \hline
2MASX J03063958+0003426 & 1949 $\pm$ 170 & 6.59 $\pm$ 0.18 & 202.43 $\pm$ 88.31 & 1.87 $\pm$ 0.02 \\
MCG+04-22-042 & 1946 $\pm$ 170 & 6.52 $\pm$ 0.06 & 95.55 $\pm$ 48.14 & 1.74 $\pm$ 0.04 \\
Mrk 110 & 2339 $\pm$ 348 & 6.45 $\pm$ 0.04 & 59.57 $\pm$ 16.68 & 1.71 $\pm$ 0.01 \\
PG 1114+445 & 5090 $\pm$ 462 & 6.42 $\pm$ 0.05 & 115.47 $\pm$ 31.05 & 1.53 $\pm$ 0.05 \\
PG 1115+407 & 1760 $\pm$ 170 & 7.18 $\pm$ 0.10 & 212.75 $\pm$ 130.91 & 2.23 $\pm$ 0.06 \\
2E1216+0700 & 2028 $\pm$ 170 & 6.48 $\pm$ 0.42 & 412.80 $\pm$ 276.98 & 2.21 $\pm$ 0.05 \\
Mrk 50 & 5540 $\pm$ 199 & 6.37 $\pm$ 0.06 & 61.53 $\pm$ 40.96 & 1.79 $\pm$ 0.03 \\
Was 61 & 1591 $\pm$ 170 & 6.56 $\pm$ 0.11 & 95.86 $\pm$ 42.68 & 2.03 $\pm$ 0.03 \\
PG 1352+183 & 4458 $\pm$ 829 & 6.45 $\pm$ 0.36 & 179.48 $\pm$ 116.27 & 1.92 $\pm$ 0.06 \\
Mrk 464 & 6188 $\pm$ 575 & 6.43 $\pm$ 0.18 & 192.70 $\pm$ 117.88 & 1.54 $\pm$ 0.06 \\
PG 1415+451 & 2850 $\pm$ 170 & 6.67 $\pm$ 0.07 & 163.30 $\pm$ 92.51 & 1.97 $\pm$ 0.05 \\
NGC 5683 & 5273 $\pm$ 313 & 6.41 $\pm$ 0.20 & 514.44 $\pm$ 336.87 & 1.74 $\pm$ 0.15 \\
Mrk 290 & 4202 $\pm$ 170 & 6.37 $\pm$ 0.06 & 74.51 $\pm$ 35.30 & 1.54 $\pm$ 0.02 \\
Mrk 493 & 988 $\pm$ 170 & 6.86 $\pm$ 0.19 & 239.25 $\pm$ 155.68 & 2.10 $\pm$ 0.05 \\
\hline
\end{tabular}
\end{center}
\end{table}
\section{Results and discussion}
\begin{figure}[t]
\begin{center}
\includegraphics[width=7.5cm]{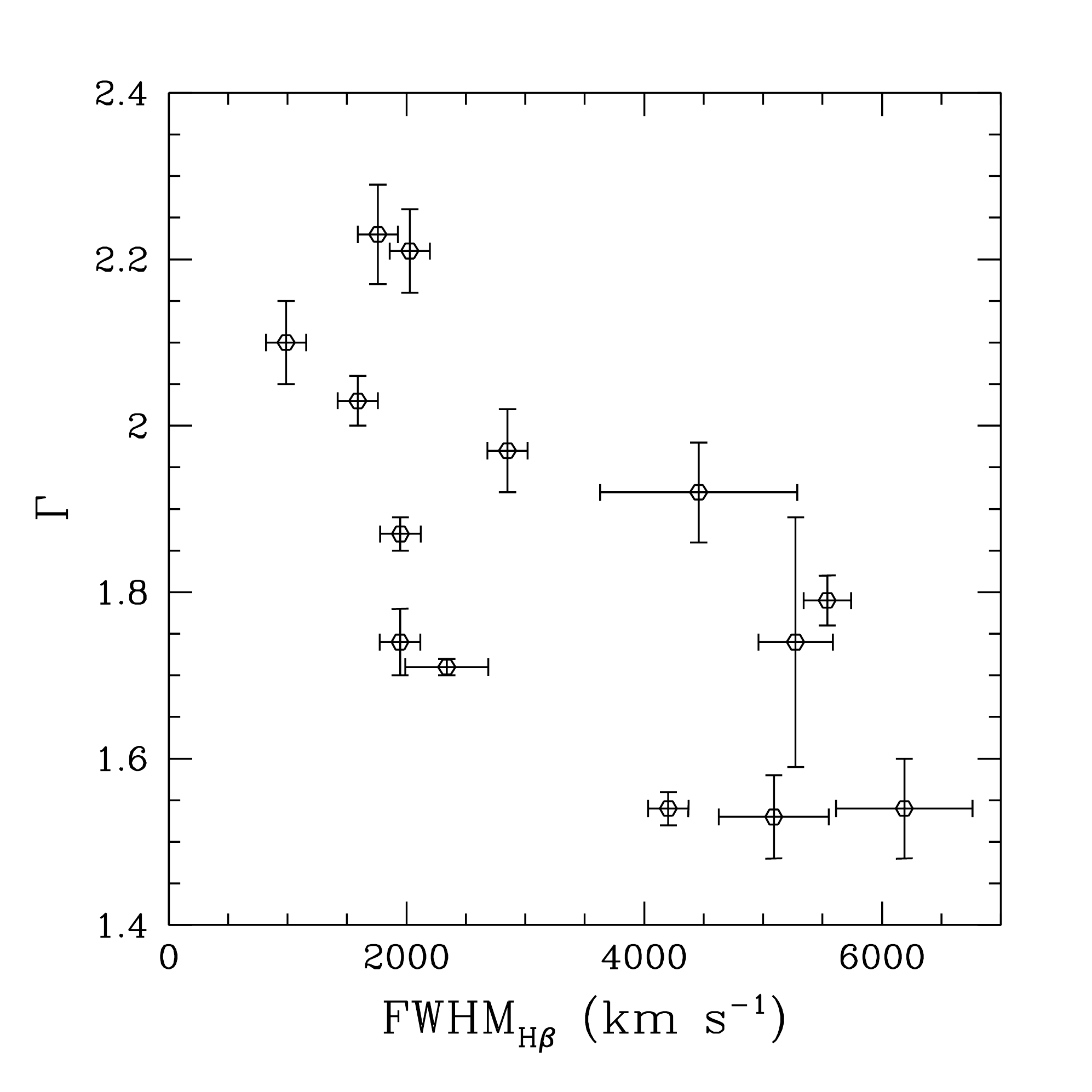}
\includegraphics[width=7.5cm]{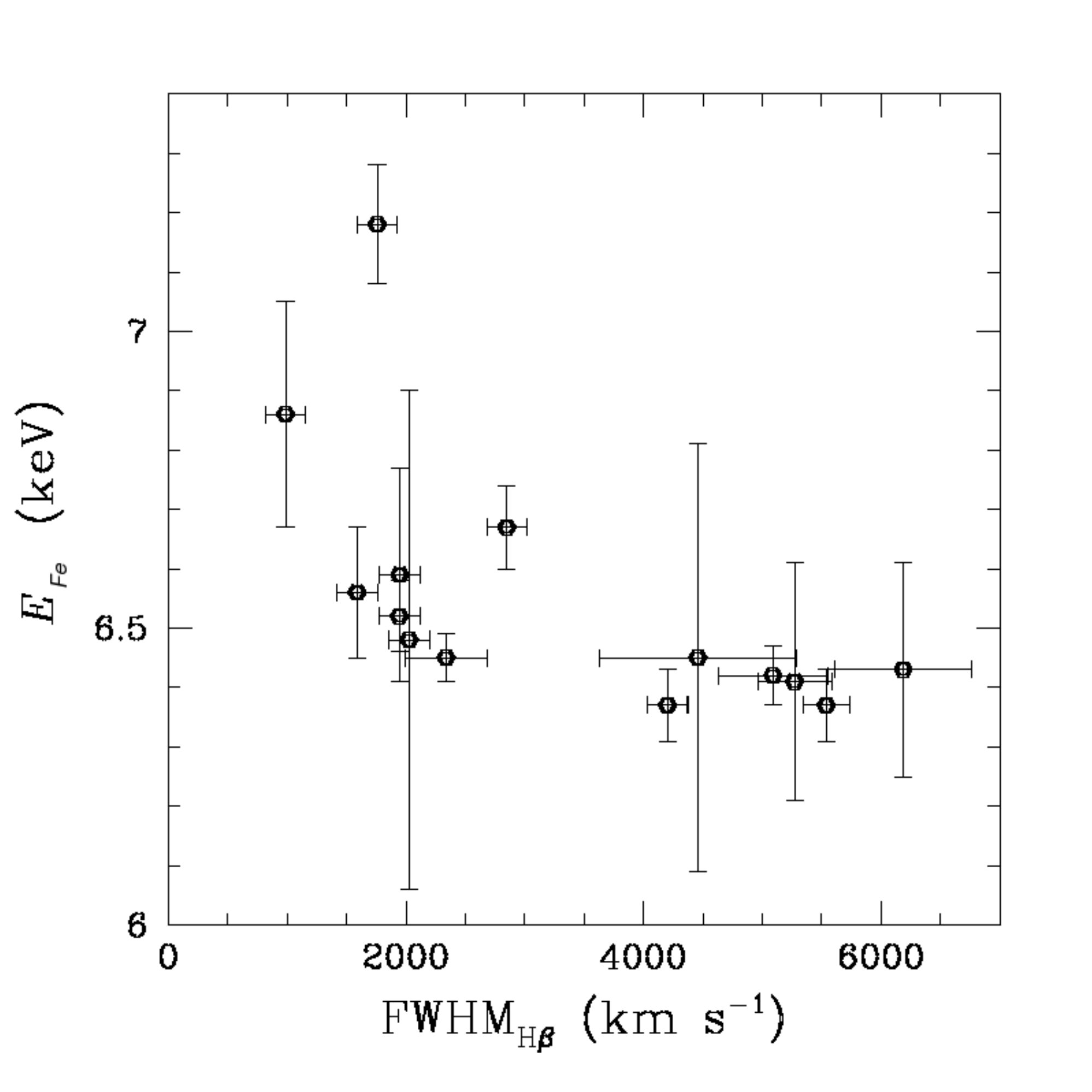}
\caption{Hard X-ray power law continuum photon index $\Gamma$ (left panel) and the Fe line central energy (right panel) plotted as functions of FWHM$_{{\rm H}\beta}$. Notice the smooth increase of the power law slope, while moving from the domain of BLS1s towards NLS1s, as opposed to the sharp raise observed in the Fe emission energy, suggesting that the BLS1s emission involves reflection from a neutral medium. \label{f05}}
\end{center}
\end{figure}

The purpose of our analysis is to combine optical and X-ray spectroscopic observations in order to explore the physics in the BLR environment of AGNs. In Tab.~\ref{t02} we collect our estimates of some fundamental parameters, which we extracted from the optical and X-ray spectra. The most notable effects, which we also illustrate in Fig.~\ref{f05}, involve the relationships of the H$\beta$ FWHM with the Fe emission line central energies and the hard X-ray power law photon index. As it was already noted \cite{Shemmer08}, we detect a clear increase in the hard X-ray power law slope moving from broad to narrow line emitters. On the other hand, reflection of the hard X-ray continuum very close to the central source originates a blend of Fe lines, with a resulting energy that depends on the ionization degree of the medium \cite{Ross99}. The most prominent contribution to this component descends from the Fe~$K\alpha$ emission of the varously ionized Fe atoms. Its energy shows a steep raise in the domain of NLS1s, indicating a dominant role of ionized reflectors in these objects. Actually, indications of reflection within a significantly ionized medium in NLS1s have been previously found (see \cite{Romano02}). In this work, however, we used the Fe line central energy to investigate the properties of the reflecting medium within a sample of targets including both narrow and broad line emitters. The results that we achieved so far suggest that, while in the range of very broad line emitting sources we observe radiation coming from a neutral component of the medium, the degree of ionization averagely increases when moving towards objects characterized by narrower emission line profiles, with a sharp raise in NLS1s.

The detection of an averagely higher ionization degree among NLS1s, with respect to broad line emitting objects, together with the different properties of the hard X-ray continuum, supports the conclusions anticipated by the BP studies of large samples of type 1 AGN optical spectra. Indeed, the high temperatures or, more generally, the stronger effects of the ionizing radiation field on the BLR plasma detected by the method are in appreciable agreement with the results observed in the X-ray domain, thus reinforcing the highly accreting low mass black hole hypothesis supported by that technique. We should note, however, that none of the properties under examination here shows a clear break distinguishing the targets simply on the basis of the width of their emission line profiles. Indeed, the complex balance of processes, resulting in the formation of the observed optical line profiles, may well affect our interpretation of the intrinsic source properties, introducing some degree of uncertainty. In La Mura et al. \cite{LaMura09}, the question of what kind of uncertainties might be expected in the determination of the central engine physical properties by means of emission line profile analysis was addressed exploiting a complex BLR kinematical model, developed to account for geometric effects, such as source flattening and inclination. By reproducing the whole shape of the emission line profiles, it was concluded that, although geometry can be expected to play a significant role in the formation of the observed spectra, it is not a critical issue for NLS1s, with respect to the remaining objects.

From the results of this work, we cannot yet draw any conclusive remark concerning many aspects of the nature of NLS1s, but we notice that the use of multiple frequency observations led us to identify consistent results from very different investigation techniques. The contribution of this analysis, as soon as it will be possible to study the whole available sample and to perform further observations, however, shall certainly lead to a better view of the intrinsic properties of these intriguing objects and to a more detailed interpretation of the innermost structure of type 1 AGNs in general.\\

We gratefully aknowledge the LOC and SOC for the invitation and for the opportunity to present this contribution. We also thank the referee for providing useful comments to develop this project.

\end{document}